\pdfoutput=1
\documentclass
[aps,prd,nofootinbib,
preprint,
superscriptaddress,eqsecnum,
]{revtex4-1}
\usepackage{graphicx,subfigure,latexsym}
\usepackage{amsmath,amssymb,amsfonts}
\usepackage{bm}
\usepackage{hyperref}
\usepackage{ulem}
\usepackage{color}
\usepackage{lineno}
\usepackage{soul}
\usepackage[]{qrcode}
\usepackage{url}

\graphicspath{{./plots/}}

\def \be {\begin{equation}}
\def \ee {\end{equation}}
\def \bes {\begin{subequations}}
\def \ees {\end{subequations}}
\def \pd {\partial}
\def \eq {
}

\def \sA {\s_{A}}
\def\muA {\mu_{A}}
\def \kp {k_{\rm peak}}
\def \kh {k_{h}}

\def\argutk {k,t}
\def \<{\langle}
\def \>{\rangle}
\def \+{\dagger}
\def \({\left(}
\def \){\right)}
\def \[{\left[}
\def \]{\right]}
\def \a {\alpha}
\def \b {\beta}

\def \g {\gamma}
\def \k {\kappa}

\def \s {\sigma}

\def \vx {\bm{x}}

\def \vr {\bm r}

\def \vj {\bm{j}}
\def \vA {\bm{A}}

\def \vL {\bm{L}}
\def \vB {\bm{B}}

\def \vE {\bm{E}}
\def \vL {\bm{L}}

\def \vW{\bm{W}}
\def \vT {\bm{T}}
\def \vP {\bm{P}}
\def \vv {\bm{v}}

\def \vX {\bm{X}}

\def \anom {\rm{A}}
\def \EM {\rm{EM}}

\def \RHIC {\rm{RHIC}}
\def \LHC {\rm{LHC}}
\def \fm {\rm{fm}}

\def \Hopf {\rm Hopf}
\def \peak {\rm p}
\def \CME {\rm {CME}}

\def\tg{\tilde{g}}

\begin{document}

%
%
%
%

\title{Self-similar inverse cascade of magnetic helicity\\ driven by the chiral anomaly}
\author{Yuji~Hirono}
\affiliation{Department of Physics and Astronomy, Stony Brook University, Stony Brook,
 New York 11794-3800, USA}
\author{Dmitri~E.~Kharzeev}
\affiliation{Department of Physics and Astronomy, Stony Brook University, Stony Brook,
 New York 11794-3800, USA}
\affiliation{Department of Physics, Brookhaven National Laboratory, Upton, New York 11973-5000}
\affiliation{RIKEN-BNL Research Center, Brookhaven National Laboratory, Upton, New York 11973-5000}
\author{Yi Yin}
\affiliation{Department of Physics,
Brookhaven National Laboratory, Upton, New York 11973-5000}

\date{\today}

\begin{abstract}
For systems with charged chiral fermions, the imbalance of chirality in the presence of magnetic field generates an electric current - this is the Chiral Magnetic Effect (CME). We study the dynamical real-time evolution of electromagnetic fields coupled by the anomaly to  the chiral charge density and the CME current by solving the Maxwell-Chern-Simons equations. We find that the CME induces the inverse cascade of magnetic helicity towards the large distances, and that at late times this cascade becomes self-similar, with universal exponents. We also find that in terms of gauge field topology the inverse cascade represents the transition from linked electric and magnetic fields (Hopfions) to the knotted configuration of magnetic field (Chandrasekhar-Kendall states). The magnetic reconnections are accompanied by the pulses of the CME current directed along the magnetic field lines. We devise an experimental signature of these phenomena in heavy ion collisions, and speculate about implications for condensed matter systems. 
\end{abstract}

\maketitle

\section{Introduction
\label{sec:intro}
}

The anomaly-induced transport of charge in systems with chiral fermions has attracted a significant interest recently. This interest stems from the possibility to study a new kind of a macroscopic quantum dynamics. 
While the macroscopic manifestations of quantum mechanics are well known (for example, superfluids, superconductors and Bose-Einstein condensates), so far they have been mostly limited to systems with broken symmetries characterized by a local order parameter, e.g. the density of Cooper pairs in superconductors. 
The effects induced by quantum anomalies in systems with chiral fermions are of different nature. 
\vskip0.3cm

Let us consider as an example the Chiral Magnetic Effect (CME) in systems with charged chiral fermions -- the generation of electric current in an external magnetic field
induced by the chirality imbalance \cite{Fukushima:2008xe}, see Refs.~\cite{Kharzeev:2013ffa,Kharzeev:2012ph,Liao:2014ava,Miransky:2015ava,Kharzeev:2015kna} for recent reviews and references. 
In this case, no symmetry has to be broken, and the system is in its normal state. However the chirality imbalance is linked by the Atiyah-Singer theorem to the non-trivial global topology of the gauge field. Since the global topology of the gauge field cannot be determined by a local measurement, there is no corresponding local order parameter, and we deal with ``topological order". 

\vskip0.3cm

This has very interesting implications for the real-time dynamics of a system composed by charged chiral fermions and a dynamical electromagnetic field. Indeed, let us initialize the system by creating a lump of chirality imbalance localized within a magnetic flux that forms a closed loop, see Fig. \ref{fig:untwisted} . Magnetic field will induce the CME current flowing along the lines of magnetic field $\vB$ (note that this effect is absent in Maxwell electromagnetism). Because the vector CME current acts as a source for the magnetic field, the current flowing along $\vB$ will twist the magnetic flux (see Fig. \ref{fig:twisted} ) and induce a non-zero expectation value for the {\it magnetic helicity} known since Gauss's work in XIX century and introduced in magnetohydrodynamics by Woltjer \cite{Woltjer} and Moffatt \cite{Moffatt}, see also \cite{ArnoldKhesin}: 
\be
\label{mag_hel}
h_m \equiv \int d^3 x\ \vA \cdot \vB \, ,
\ee
where $\vA$ is the vector gauge potential.
Magnetic helicity is a topological invariant (Chern-Simons three-form) characterizing the global topology of the gauge field. It is mathematically related to the knot invariant, and measures the chirality of the knot formed by the lines of magnetic field. Because of this, the generation of magnetic helicity will create the chiral knot out of the closed loop of magnetic flux -- so the topology of magnetic flux will change. In this paper we will quantify this statement, and study how the topology of magnetic flux changes in real time. We will find that as a consequence of chiral anomaly and the CME, the magnetic field evolves to the self-linked Chandrasekhar - Kendall states (see Fig. \ref{fig:CK} ). 
During the evolution, the size of the knot of magnetic flux increases. Moreover, at late times this evolution becomes self-similar, and is characterized by universal exponents. 
\vskip0.3cm
The evolution of magnetic helicity has been studied previously in the framework of the Maxwell-Chern-Simons theory in Refs.~\cite{McLerran:2013hla,Tuchin:2014iua,Manuel:2015zpa}
(see Ref.~\cite{Jackiw:1999bd} for study with Maxwell theory). 
The anomaly-driven inverse cascade is discussed in Refs.~\cite{Joyce:1997uy,PhysRevLett.108.031301,Boyarsky:2015faa}.
However the self-similar evolution of magnetic helicity has not been reported in these papers. 
The closest to our present study is the paper~\cite{Tashiro:2012mf} by Tashiro, Vachaspati and Vilenkin
, 
where a simplified version of the anomalous magneto-hydrodynamic equation has been applied to cosmic magnetic fields. 
The authors  found the power law decay (in terms of conformal time) of the chiral chemical potential at the late stage of evolution (see also Ref.~\cite{PhysRevLett.108.031301}). 
\vskip0.3cm

We extend the previous studies by elucidating the topology of magnetic flux throughout the evolution of magnetic helicity. This is made possible by the use of the  eigenfunctions of the curl operator in a spherically symmetric domain. Previous studies~\cite{PhysRevLett.108.031301,Manuel:2015zpa} have used the eigenfunctions of curl operator in a free-space, i.e. the polarized plane waves. In our treatment we can track the magnetic reconnections that transfer helicity from linked to self-linked configurations of magnetic flux. We also identify the final state of the system as the Chandrasekhar-Kendall state that minimizes the magnetic energy at fixed helicity. 
 
\vskip0.3cm
This paper is organized as follows. In Sec.~\ref{sec:topology} we describe the topology of magnetic flux and describe the corresponding solutions. 
In Sec.~\ref{sec:qualitative},
we introduce magnetic helicity spectrum and present a qualitative picture of the inverse cascade of magnetic helicity and the role played by anomaly. 
In Sec.~\ref{sec:CSM},
we introduce the Maxwell-Chern-Simons equations which we will use to study the evolution of magnetic helicity and axial charge density. 
The results of evolution are presented in Sec.~\ref{sec:evolution}.
In Sec.~\ref{sec:HIC},
we discuss the relevance of our findings for heavy-ion collision experiment. 
We conclude and discuss possible extensions of the current work in Sec.~\ref{sec:conclusion}.

\section{The chiral anomaly and topology of magnetic flux}\label{sec:topology}

Consider a link ${\cal K}$ of $N$ knots of magnetic field with fluxes $\phi_i$. The corresponding magnetic helicity (\ref{mag_hel}) of this link is given by \cite{Moffatt,BF84,MR92,Berger}
\be
\label{mag_top}
h_m ({\cal K}) = \sum_{i=1}^N \phi_i^2\ {\cal S}_i + 2 \sum_{i,j} \phi_i \phi_j\ {\cal L}_{ij}\, , 
\ee
where ${\cal S}_i$ is the C\u{a}lug\u{a}reanu-White self-linking number, and ${\cal L}_{ij}$ is the Gauss linking number\footnote{The same formula applies to the helicity of vortex flows, with the substitution of gauge potential $\vA$ by the velocity field $\vv$, magnetic field $\vB$ by vorticity ${\bm \omega} = {\bm \nabla} \times \vv$, and the flux $\phi_i$ by the circulation $\kappa_i$.}. The linking numbers in (\ref{mag_top}) do not always detect the topology of the link; the development of the appropriate knot invariants is a very active area of modern mathematics. The link between the Jones invariant of the knot and Chern-Simons theory has been uncovered by Witten \cite{Witten:1988hf}. The recent progress includes the HOMFLY knot polynomials, Vasiliev invariants, Khovanov and Heegaard-Floer homologies, but the ultimate solution is still lacking. In view of this, we will base our discussion on formula (\ref{mag_top}). 

\begin{figure}
\centering
\subfigure[]
{
   \label{fig:untwisted} 
        \includegraphics[width=7cm]{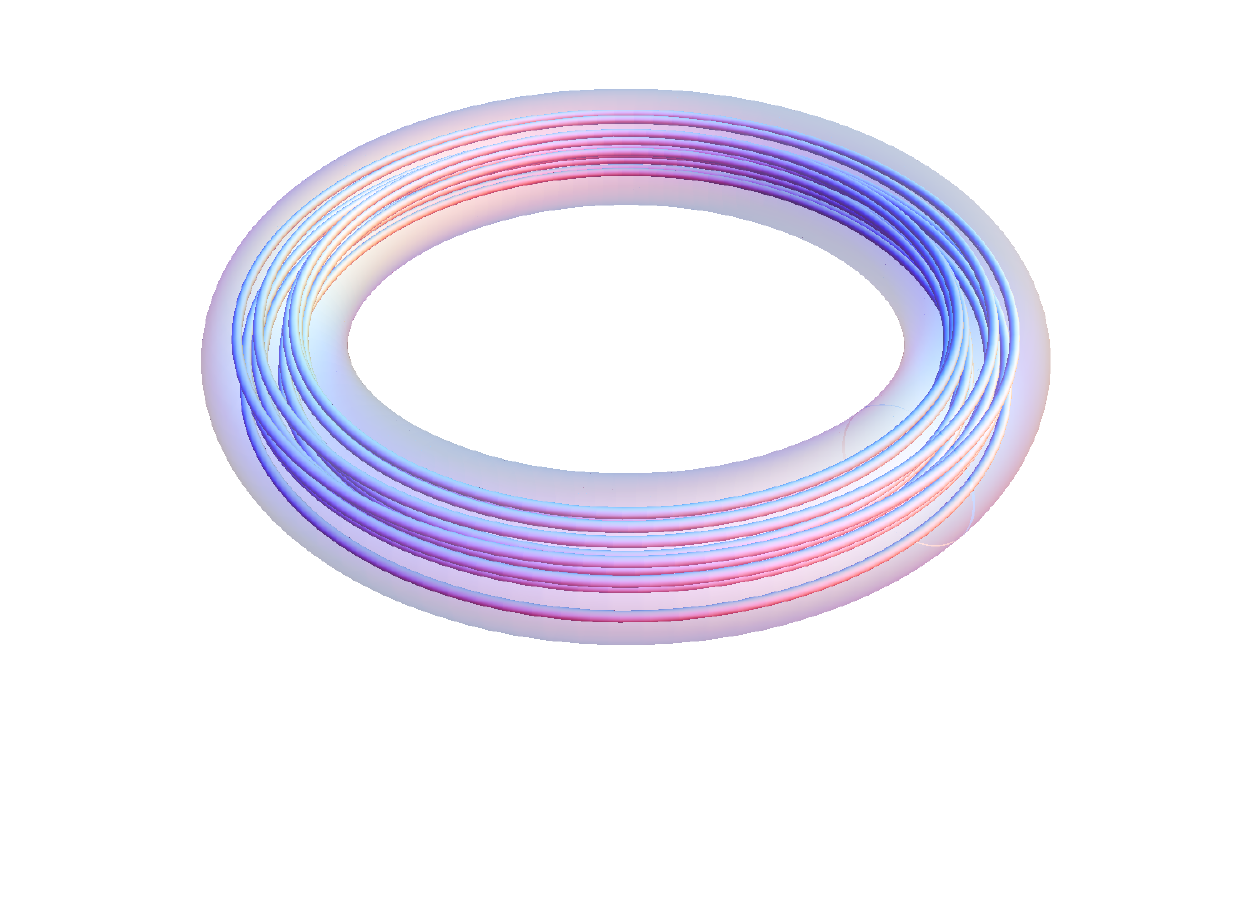}
} 
\subfigure[]
{
   \label{fig:twisted} 
        \includegraphics[width=7cm]{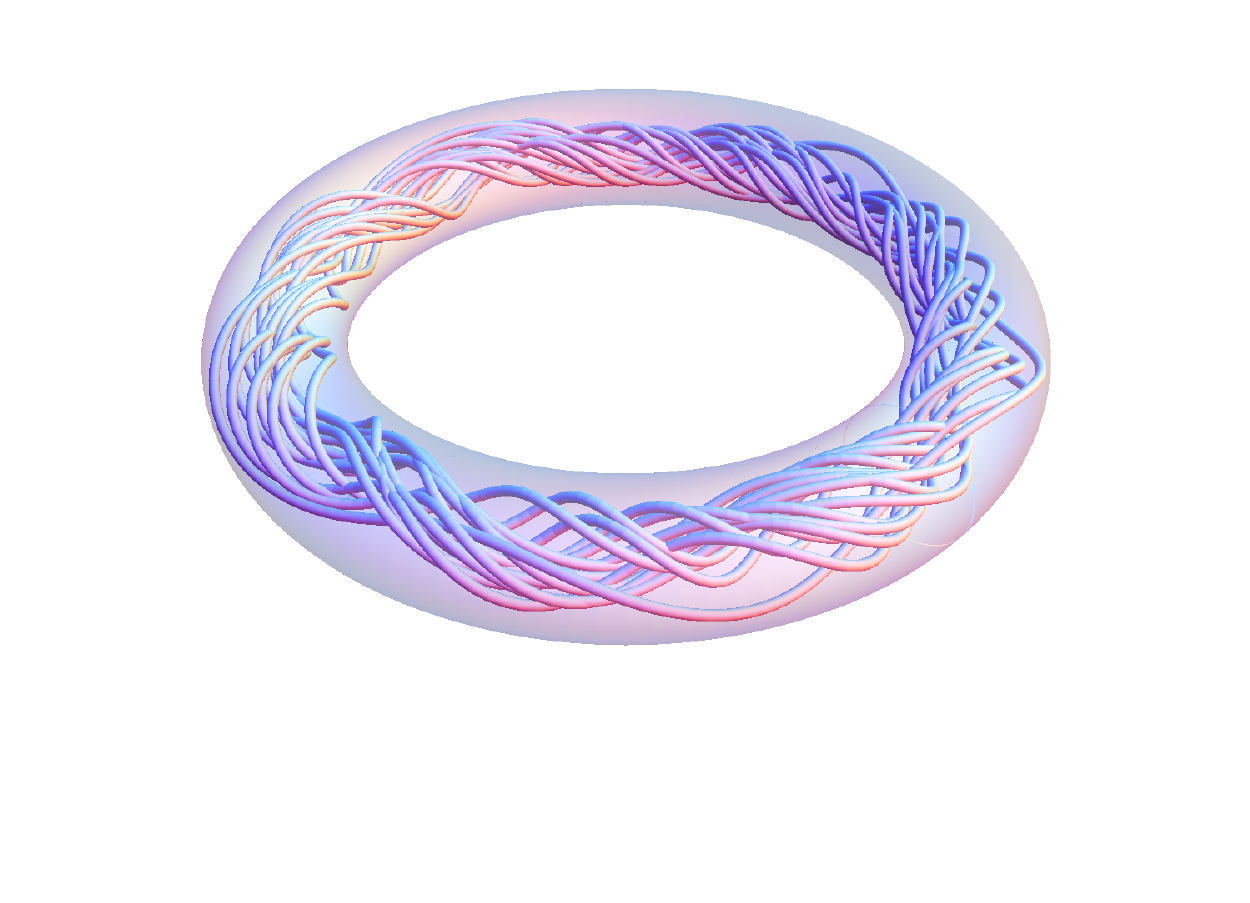}
} 
\subfigure[]
{
   \label{fig:trefoil} 
        \includegraphics[width=7cm]{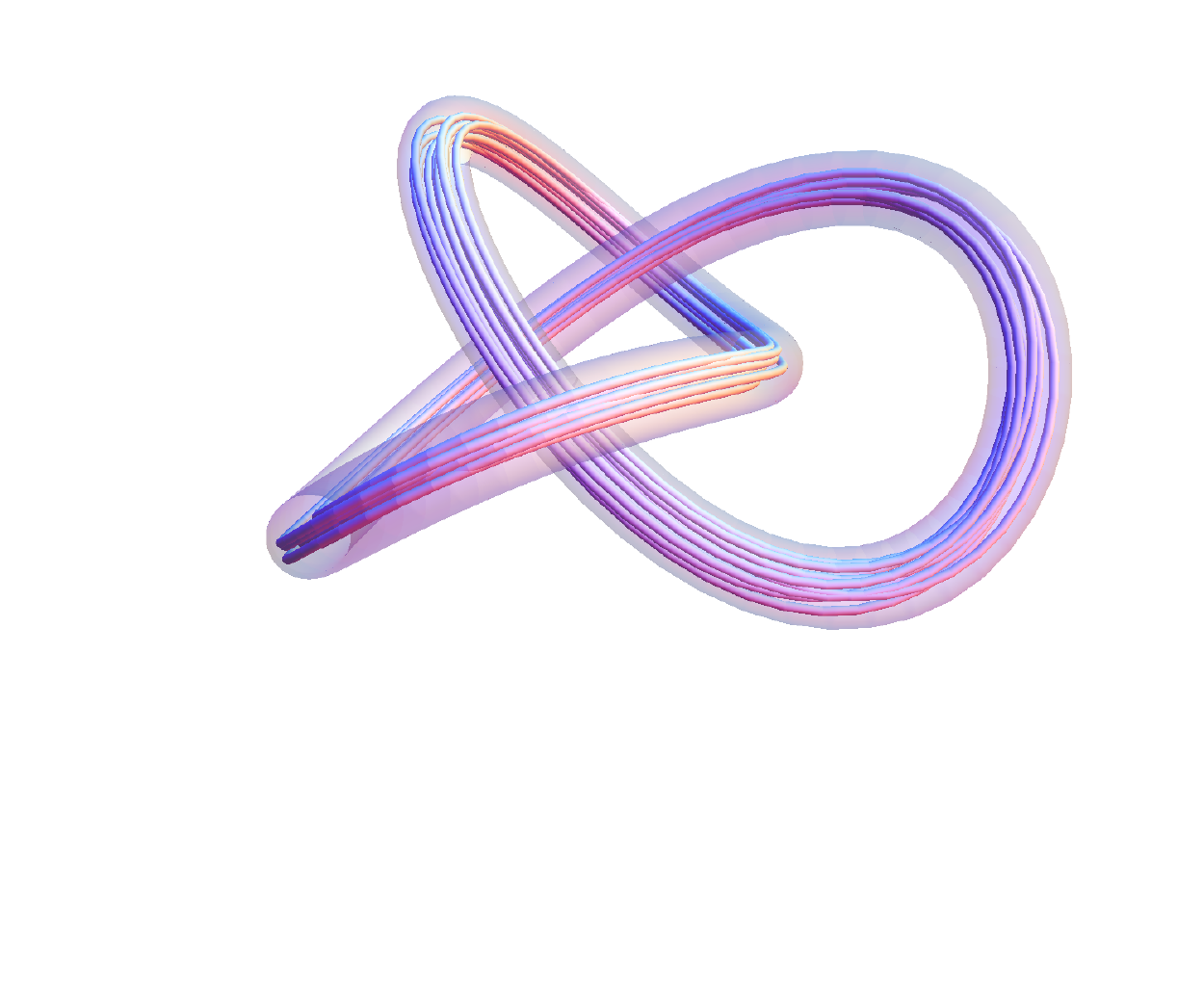}
} 
\subfigure[]
{
   \label{fig:CK} 
        \includegraphics[width=7cm]{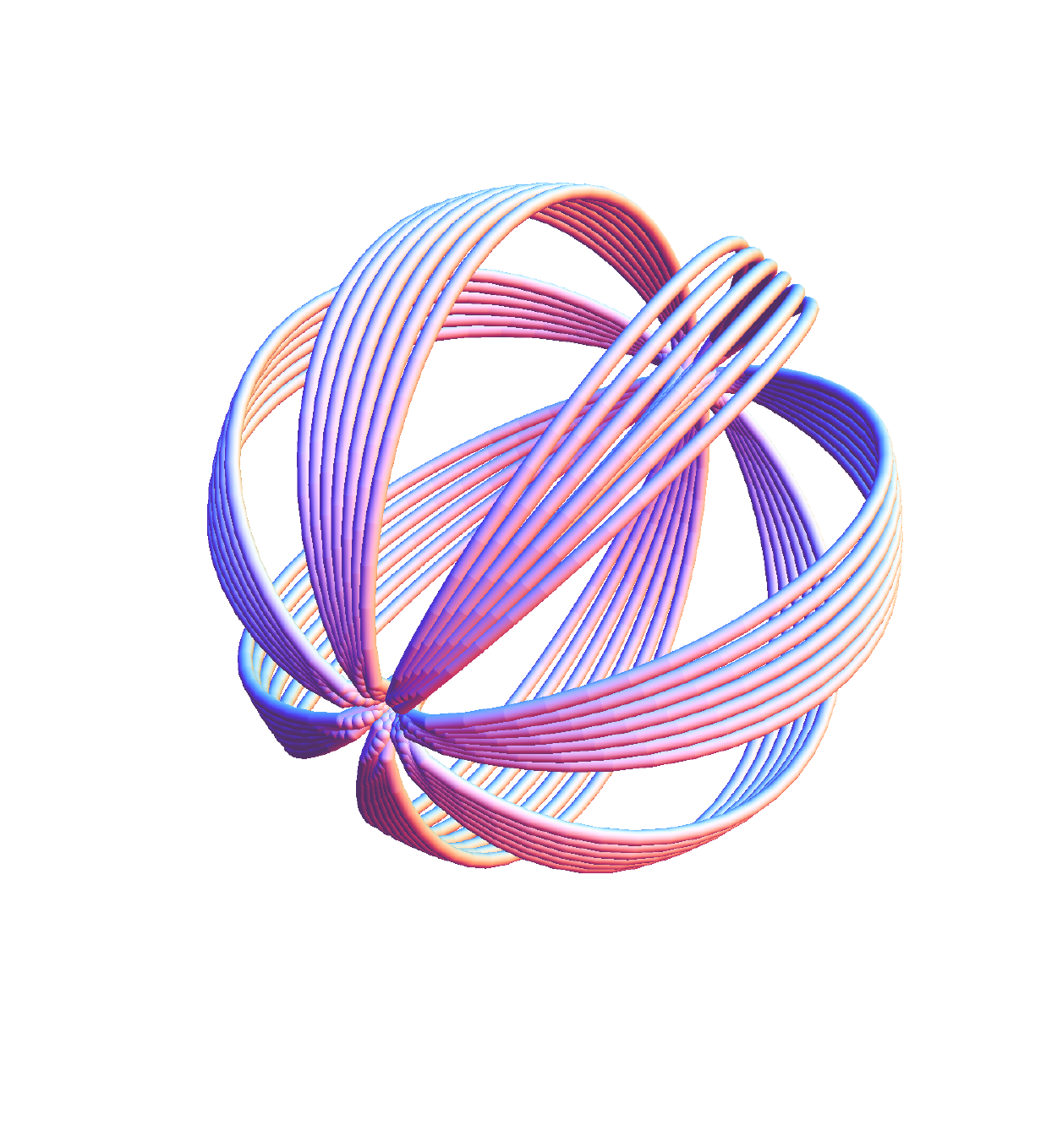}
} 
\caption{
\label{fig:topology}
The topology of Abelian magnetic flux: (a) upper left -- untwisted loop; 
(b) upper right -- twisted magnetic flux; 
(c) lower left -- the self-linked magnetic flux (trefoil knot shown); 
(d) lower right -- the self-linked  Chandrasekhar-Kendall state.
}
\end{figure}

\vskip0.3cm
In MagnetoHydroDynamics (MHD), the lines of magnetic field are ``frozen" into the fluid, and so the magnetic helicity (\ref{mag_top}) is conserved. Moreover, in the absence of dissipation the reconnections of magnetic field are absent, and so the topology of the knotted configuration is preserved as well -- so the two terms in (\ref{mag_top}) are conserved separately. As we will now discuss, in fluids with charged chiral fermions the situation changes dramatically due to the presence of chiral anomaly. Indeed, the anomaly relation 
\be
\label{anomaly0}
\pd_{\mu}j^{\mu}_{A} = C_{\anom} \vE\cdot\vB\, 
\ee
describes the generation of chirality by electric $\vE$ and magnetic $\vB$ fields in a topologically non-trivial configuration characterized by Chern-Pontryagin number density $\vE\cdot\vB$. The anomaly coefficient $C_{\anom}$ for the case of QCD plasma containing $N_c$ colors and $N_f$ flavors of quarks is given by $C_{\anom} = N_{c}C_{\EM}e^2/2\pi^2$, with $C_{\EM} = \sum_{f} q^2_f $. The Chern-Pontryagin number is easily seen to be equal to the time derivative of magnetic helicity:
\be
\int d^3 x\ \vE\cdot\vB\ = - \frac{1}{2}\ \frac{\partial h_m}{\partial t}\, .
\ee
This means that when chirality of the fermions is changed, this change is accompanied by the change of magnetic helicity, implying the reconnection of magnetic flux. Reconnections of magnetic flux in particular can cause transitions between the self-linked (see Figs. \ref{fig:trefoil}  and \ref{fig:CK}) and linked configurations of magnetic field described by the first and the second terms in (\ref{mag_top}) respectively. 
Below we will show that such transitions indeed happen as a consequence of the anomaly, and the system evolves towards the state in which magnetic flux is self-linked, i.e. the entire magnetic helicity is given by the first term of (\ref{mag_top}).
\vskip0.3cm
Before proceeding to the calculations, let us discuss the possible topologies of magnetic flux. The Maxwell equations in free space allow for simple solutions with non-zero magnetic helicity -- these solutions are just circularly polarized plane waves. This is intuitively clear since magnetic helicity is parity-odd, and left- and right-circularly polarized waves are the simplest $P$-odd states of electromagnetic field.  Since we are interested in describing the plasma of a final extent in space, we have however to impose the boundary conditions on electromagnetic field. In this case the solutions of Maxwell equations are given by Hopfions \cite{Hopf} -- configurations in which the loops of magnetic and electric fields are linked.  
\vskip0.3cm


On the other hand, the dynamics of electromagnetic fields in the presence of chiral anomaly is described by Maxwell-Chern-Simons theory. In particular, the chiral imbalance between the left- and right-handed fermions quantified by the chiral chemical potential $\muA$ leads to the generation of electric CME current \cite{Fukushima:2008xe} along the magnetic field:
\be
\label{CME}
\vj_{\CME}= C_{\anom}\ \muA\ \vB = \sA \vB\, ,  
\ee
where we introduced the ``chiral magnetic conductivity" $\sA$ to allow for the frequency dependence \cite{Kharzeev:2009pj}. Unlike the usual Ohmic current, the CME current is topologically protected and hence non-dissipative. Hence at late times when Ohmic currents have already dissipated away, the r.h.s. of the Maxwell equation $\nabla\times\vB = {\bm j}$ will contain only the CME current and will thus acquire the form
\be\label{curlB}
\nabla\times\vB = \sA\ \vB\, .
\ee 
The solutions of (\ref{curlB}) have been found independently\footnote{The Editor of their paper \cite{CK} wrote: ``The results in this paper were derived independently by the two authors, and they agreed to write it as one.".}  by Chandrasekhar and Kendall \cite{CK}; we will refer to them as CK states, and illustrate their structure in Fig.~\ref{fig:CK} ). 
It has been found by Woltjer \cite{Woltjer} that these ``force-free" configurations of magnetic field that obey (\ref{curlB}) minimize the total magnetic energy 
\be
{\cal E}_{M}\equiv  \frac{1}{2}\int d^3 x\ \vB^2
\ee
at a given magnetic helicity (\ref{mag_hel}). 
We thus expect that the CME currents will lead to the transition from Hopfion states to CK states at late times, once the Ohmic currents have dissipated. We will see below that explicit computations indeed yield this result. 

\section{Inverse cascade of magnetic helicity driven by anomaly
\label{sec:qualitative}
}

\subsection{Magnetic helicity spectrum and Chardrasekhar-Kendal (CK) states
\label{sec:CK_states}
}

To discuss the spatial distribution and the inverse cascade of magnetic helicity, let us specify the structure of CK states Ref.~\cite{CK} that are defined as eigen-functions of the curl operator, see (\ref{curlB}). 
In a free space, a CK state is nothing but a circularly polarized plane wave.
In this work however,
we are interested in EM field in a finite closed system. Let us thus consider CK states $W^{\pm}_{lm}(\vx;k)$ in a spherical domain:
\be
\nabla \times \vW^{\pm}_{lm}(\vx;k)
= \pm k \vW^{\pm}_{lm}(\vx;k)\, ,
\qquad 
\nabla\cdot \vW^{\pm}_{lm} (\vx;k)=0\, ,
\ee 
where $l=0, 1, \ldots, m=-l, -l+1, \ldots l$.
The explicit expressions for $W^{\pm}_{lm}(\vx;k)$ in terms of spherical
harmonics and spherical Bessel functions are given in Appendix.~\ref{sec:CK}.
For the purposes of the present discussion,
we only need to keep in mind that they form a complete basis for any divergence-less vector and satisfy the orthogonality relation:
\be
\int d^3x\, \vW^{a}_{lm}(\vx;k)\cdot \vW^{b}_{l'm'}(\vx;k')
= \frac{\pi}{k^2}\delta(k-k')\delta_{ll'}\delta_{mm'}\delta_{ab}
\, ,
\qquad
a,b=+,-\, .
\ee
\vskip0.3cm

Let us now expand magnetic field $\vB$ in terms of CK states $\vW^{\pm}_{lm}(\vx;k)$:
\be
\label{BinW}
\vB(\vx, t) = \sum_{l, m}\int^{\infty}_{0}\frac{dk}{\pi} k^2 
\[ \a^{+}_{lm}(\argutk) \vW^{+}_{lm}(
\vx;k)+\a^{-}_{lm}(\argutk) \vW^{-}_{lm}(\vx;k)\]\, .
\ee
Using $\nabla \times \vA=\vB$, 
we also expand $\vA(\vx, t)$ as\footnote{We choose the gauge
  $\nabla\cdot \vA =0$ ; the magnetic helicity $h_{m}$ is gauge invariant.}:
\be
\vA(\vx, t) = \sum_{l,m}\int^{\infty}_{0}\frac{dk}{\pi} k
\[ \a^{+}_{lm}(\argutk) \vW^{+}_{lm}(
\vx;k)-\a^{-}_{lm}(\argutk) \vW^{-}_{lm}(\vx;k)\]\, . 
\ee
Consequently, 
the magnetic helicity $h_{m}$ 
\be
h_{m}\equiv \int d^3x\ \vA\cdot \vB\, ,
\ee
 is now given by:
\be
\label{h_and_g}
h_{m}(t)
= \int^{\infty}_{0}\frac{dk}{\pi}\ k g(k,t)\, , 
\qquad
g(k,t)\equiv
g_{+}(k, t)- g_{-}(k, t)\, . 
\ee
Here we have defined the \textit{magnetic helicity spectrum} $g(\argutk)$; the functions  $g_{\pm}(\argutk)$ describe the relative weight of a single CK state $W^{\pm}_{lm}(\vx;k)$ with a definite helicity:
\be
\label{g_def}
g_{\pm}(k, t)
\equiv \sum_{l, m} |\a^{\pm}_{lm}(k, t)|^2\, . 
\ee
The energy of magnetic field can be related to $g_{\pm}(k)$:
\be
\label{B_E}
{\cal E}_{M}
\equiv \int d^3x \, \frac{1}{2}\vB^2
= \frac{1}{2}\int ^{\infty}_{0}\frac{dk}{\pi}
k^2\[g_{+}(k, t)+g_{-}(k, t)\]\, . 
\ee
Comparing \eq\eqref{h_and_g} with \eq\eqref{B_E},
we find that the energy cost for a CK state $W^{\pm}(\vx, k)$ to carry one unit of helcity is $k$.

\subsection{Qualitative picture of the anomaly-driven inverse cascade of magnetic helicity
\label{sec:energy}
}

We now ready to discuss the qualitative picture of the inverse cascade driven by anomaly. 
Let us first define the {\it fermionic helicity}:
\be
\label{hF_def}
h_{F}\equiv C^{-1}_{\anom}\int d^{3}x\,\ n_{A}\, ,
\ee
where $n_A = j^0_A$ is the density of axial charge. 
From the anomaly equation \eq\eqref{anomaly0} and the Maxwell equations,
the total helicity of the system $h_{0}$ is conserved: 
\be
\label{h0}
h_{0}\equiv 
h_{m}+h_{F}
=\text{const}\, . 
\ee
Therefore, the 
system will tend to minimize the energy cost at fixed helicity. 
From the definition of ``fermionic helicity'' \eq\eqref{hF_def} and
linearized equation of state $n_{A}=\chi \muA$ where $\chi$ is the susceptibility, 
we observe that the energy per fermonic helicity is $\sA$.
On the other hand, 
as we discussed in Sec. \ref{sec:CK_states},
the energy per magnetic helicity for a single CK mode is $k$. 
Therefore for a hard (positive, i.e. of right circular polarization) CK mode $k>\sA$ (where without a loss of generality, we take $\sA$ to be positive as well), carrying helicity by chiral fermions is energetically favorable and consequently the helicity will be transferred from hard CK modes to chiral fermions. In contrast, the soft CK modes with $k<\sA$ are energetically favorable compared to chiral fermions, and fermionic helicity will be transferred to the soft components of magnetic helicity. Because the total helicity is conserved, this transfer will deplete the value of $\sA$, and so the transfer of fermionic helicity will occur to softer and softer CK modes -- therefore, we find an inverse cascade of magnetic helicity. 
\vskip0.3cm

We are now ready to extend our discussion of individual modes to the evolution of the entire helicity spectrum
$g(\argutk)$.
For definiteness, let us assume that the total helicity of the system $h_{0}$ is positive. It is convenient to 
 introduce a characteristic energy scale $k_{h}$ associated with total helicity:
\be
\label{kh_def}
\kh 
\equiv\frac{C^{2}_{A}h_{0}}{\chi V}\, ,
\ee
with $V$ the volume. 
The quantity $k_{h}$ can be interpreted as the energy per helicity if the entire helicity $h_{0}$ is carried by chiral fermions. 
The fate of the system depends on the values of $k_{h}$ and $k_{\rm min}$, 
the lowest possible eigenvalue of a CK state allowed by the boundary conditions. 
If $k_{h}<k_{\rm min}$, 
eventually all magnetic helicity will be transferred to fermonic helicity. 
In contrast, 
if $k_{h}>k_{\rm min}$, 
the helicity will eventually be carried by magnetic fields and the configuration of magnetic field will approach a single CK state $W^{+}_{lm}(k_{\rm min})$ that minimizes the energy at a fixed helicity. 
We will confirm this scenario by a quantitative analysis in the next section. 
 
 \section{Maxwell-Chern-Simons equations
\label{sec:CSM} 
 }
 
The CME current can be described by adding the Chern-Simons term to the Maxwell theory  \cite{Kharzeev:2009fn}. Assuming that the gradients of chirality distribution are small, the resulting 
sourceless Maxwell-Chern-Simons (MCS) theory acquires the usual Maxwell form with an extra term in the current describing the CME:
\bes
\be
\label{maxwell1}
\nabla\times \vB
= \frac{\pd\vE}{\pd t}+\vj_{\EM} \, ,
\ee
\be
\label{maxwell2}
\nabla\times \vE
= - \frac{\pd \vB }{\pd t}\, .
\ee
\be
\label{maxwell3}
\nabla \cdot \vB =0\, , 
\qquad
\nabla \cdot \vE =0\, . 
\ee
\ees
Here, the electric current $\vj_{\EM}$ includes the Ohmic and CME components:
\begin{equation}
\label{j}
\vj_{\EM} = \s \vE + \s_A \vB\, ,
\end{equation}
where $\s$ is the usual electrical conductivity. Let us now take the curl of the equation \eq\eqref{maxwell1}
and use \eq\eqref{maxwell2} and \eq\eqref{maxwell3} to obtain
\bes
\label{dyn}
\be
\label{B_evo}
\s\pd_{t}\vB(t,\vx)
= \nabla^2\vB +\sA \(\nabla \times \vB \)\, .
\ee
In \eq\eqref{B_evo},
we have neglected $\pd^2_{t}\vB$ term -- 
this should be a good approximation for time scales larger than $1/\s$. 
Here, as mentioned above, 
we also neglect the spatial dependence of $n_{A}$ and relate $\mu_{A}$ to $n_{A}$ via the linearized equation of state $\mu_{A}=n_{A}/\chi$; this yields 
\be
\label{approx}
\s_{A}(t)
= \frac{C_{A}n_{A}(t)}{\chi}
\approx \frac{C_{A}}{\chi V}\int d^3x\, n_{A}(\vx, t)\, .
\ee
In accord with our assumption of small gradients of the axial density 
we will neglect the spatial component of axial current $\vj_{A}$.
The evolution of $n_{A}(t)$ is thus related to the evolution of Chern-Pontryagin density 
$\vE\cdot \vB$ by the anomaly equation \eq\eqref{anomaly0}:
\be
\label{anomaly}
\pd_{t} n_{A}(t) = \frac{C_{\anom}}{V}\int d^3x\vE\cdot \vB\, .
\ee
\ees
Eqs.~\eqref{dyn}
give the simplified version of the MCS equations that we about to solve.


\subsection{General solutions}
Once we apply the decomposition \eqref{BinW},
\eq\eqref{B_evo} becomes a differential equation describing the time dependence of $\a_{lm}(\argutk)$:
\be
\label{master_equation}
\pd_{t} 
\a^{\pm}_{lm}(\argutk)
= \s^{-1}\[-k^2\pm \sA(t) k 
\]\a^{\pm}_{lm}(\argutk)\, .
\ee
From the definition of chiral magnetic conductivity $\s_{A}=C_{A}\muA$ and \eq\eqref{h0}, 
we have
\be
\label{sA1}
\sA(t) 
= \kh
\[1 - \frac{h_{m}(t)}{h_{0}}\]\, ,
\ee
where $\kh$ is defined in \eq\eqref{kh_def}.
The solution to
\eq\eqref{master_equation} and \eq\eqref{sA1} can be obtained as
follows (see also Refs.~\cite{Tashiro:2012mf,Manuel:2015zpa}).
First, integrating \eq\eqref{master_equation}, 
we get
\be
\a^{\pm}_{lm}(\argutk)
=\a^{\pm}_{lm, I}(k)\, \exp\Bigg\{\s^{-1}\[-k^2t\pm k\,\theta(t)\]\Bigg\}\, , 
\ee
where $\a^{\pm}_{lm, I}(k)\equiv \a^{\pm}_{lm}(t=0,k)$ is determined
by the initial value of magnetic field and 
\be
\label{theta_def}
\theta(t)\equiv
\int^{t}_{0}dt'\, \sA(t')\, . 
\ee
Now, from the definition \eq\eqref{g_def} we have
\be
\label{g_sol}
g^{\pm}(\argutk)
=g^{\pm}_{I}(k)
\exp\Bigg\{2\s^{-1}\[-k^2t\pm k\,\theta(t) \]\Bigg\}\, ,
\ee
where $g^{\pm}_{I}(k)\equiv g^{\pm}_{I}(t=0, k)$ denotes the initial magnetic helicity spectrum. 
Finally,
$\theta(t)$ (and thus $\sA(t)$) will be determined from the consistency condition
\eq\eqref{sA1}:
\be
\label{sA2}
\sA(t)
= \kh 
\Bigg\{1 -
\frac{1}{h_{0}}\int^{\infty}_{0}\frac{dk}{\pi}k \[ g_{+}(\argutk)-g_{-}(\argutk)\]\Bigg\}\,
. 
\ee

Before presenting numerical solutions,
we now discuss the evolution of individual CK modes
$\a^{\pm}_{lm}(\argutk)$ as described by \eq\eqref{master_equation}. 
Without losing generality,
let us assume that $\sA(t)>0$. 
Then the negative helicity mode $\a^{-}_{lm}(\argutk)$ will decay exponentially, so let us concentrate on the evolution of the positive helicity mode $\a^{+}_{lm}(\argutk)$.
 For hard modes, i.e. the modes with momenta $k\gg \sA$, 
$\a^{+}_{lm}(\argutk)$ decays exponentially $\exp(-\s^{-1}k^2t)$, as
usual. 
However, the 
soft helicity mode $k<\sA(t)$ will grow exponentially. 
This unstable mode has been noticed before in various contexts, see Refs.~\cite{Joyce:1997uy,Akamatsu:2013pjd,Tuchin:2014iua,Manuel:2015zpa} for examples. 
The growth of soft CK modes could be anticipated from the discussion in Sec.~\ref{sec:energy}: the system tends to minimize the total energy while preserving the total helicity, and the soft CK state possesses the lowest energy at a fixed helicity. 

\section{The inverse cascade of magnetic helicity}
\label{sec:evolution}

\subsection{The initial conditions and Hopfion solutions}
\label{sec:IC}

As discussed in the previous section, 
the evolution of $\theta(t)$, $\sA(t)$ and $g(\argutk)$ can be determined once the initial
condition for the configuration of electromagnetic field is specified. Since we would like to investigate the evolution of topology of magnetic flux, we 
assume that initially the electromagnetic field with a non-zero magnetic helicity is localized at a short spatial scale 
 much shorter than
$k^{-1}_{h}$ defined in \eq\eqref{kh_def}.
We therefore take the Hopfion solution~\cite{Hopf} to vacuum Maxwell equations as the initial configuration. 
This solution carries non-zero helicity (which we assume
to be positive) that is due to the second term in (\ref{mag_top}) and a finite energy.
It may be interpreted as a soliton wave solution to the Maxwell
equations.  
For a Hopfion solution with a total initial magnetic helicity $h_{m, I}$, 
the electromagnetic field can be expressed in terms of CK states as~\cite{Hopfion_nature} (see also Ref.~\cite{Hoyos:2015bxa}):
\bes
\label{Hopf_sol}
\be
\vB_{\Hopf}(\vx,t)
= 
\sqrt{\frac{4 h_{m, I}}{3\pi}}\int^{\infty}_{0} dk k^2 e^{-k L_{\EM}}
\[( k L^2_{\EM})\vW^{+}_{11}(\vx, t)e^{-ikt}+\text{c.c.}\]\, ,
\ee
\be
\vE_{\Hopf}(\vx,t)
= 
\sqrt{\frac{4 h_{m, I}}{3\pi}}\int^{\infty}_{0} dk k^2 e^{-k L_{\EM}}
\[(-i k L^2_{\EM})\vW^{+}_{11}(\vx, t)e^{-i kt}+\text{c.c.}\]\, ,
\ee
\ees
where $L_{\EM}$ characterizes the size of the Hopfion. 
Consequently,
only modes $\a^{+}_{11}(\argutk)$ are non-vanishing (see (\ref{BinW})) and 
the initial magnetic helicity  spectrum is given by:
\be
\label{g_I}
g_{I}(k)= \frac{8\pi}{3}h_{m, I}L^{4}_{\EM}k^2e^{-2k L_{\EM}}\, . 
\ee
The peak of $g_{I}(k)$ at $k_{\peak} \equiv 1/L_{\EM}$ defines a characteristic size $L_{\EM}$ of the configuration. 

\subsection{Stages of the inverse cascade evolution}
\label{sec:stage}
We are now ready to solve the evolution equations \eq\eqref{master_equation} and \eq\eqref{sA2}.
In addition to the initial condition discussed in Sec.~\ref{sec:IC},
the evolution also depends on the dimensionless ratios $L_{\EM} k_{h}, h_{m, I}/h_{0}, \s/\kh $. 
We would like to model the situation in which initially the helicity is dominated by the contribution from the EM field, thus $h_{m}\approx h_{0}$ and $L_{\EM}>\kh^{-1}$.
To be concrete, in this subsection we present the results corresponding to the solution with 
$\(L_{\EM} k_{h}, h_{m, I}/h_{0}, \s/\kh\)=\(1/2, 0.8,0.4\)$.
We have also numerically solved \eq\eqref{sA2} with different choices of $L_{\EM} k_{h}, h_{m, I}/h_{0}, \s/\kh $; the results are qualitatively similar. 

\begin{figure}
\centering
        \includegraphics[width=0.6\textwidth]{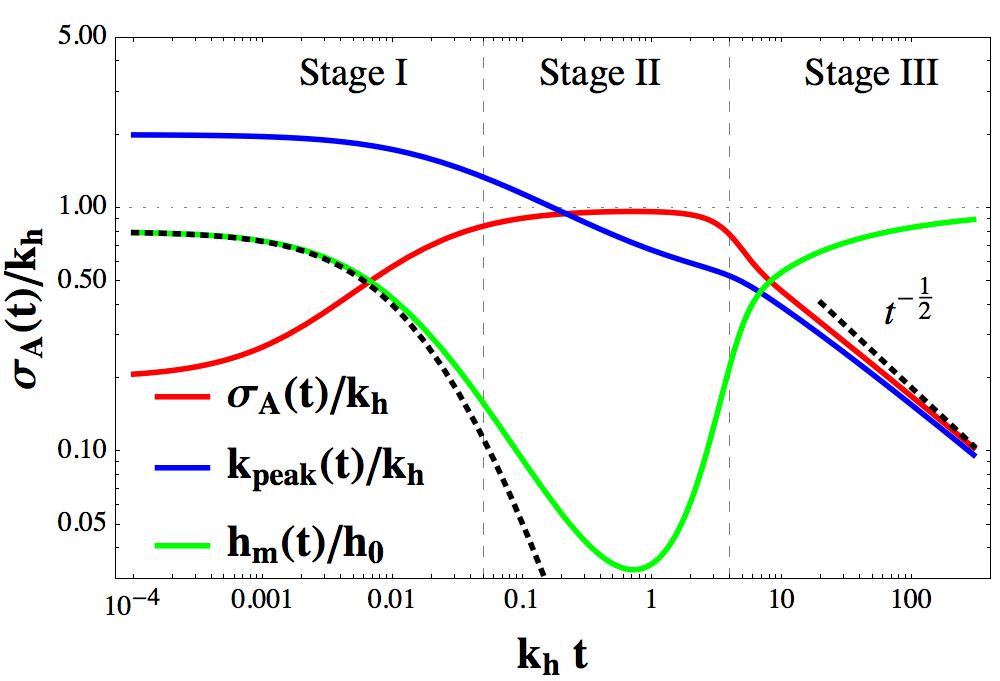}
\caption{
\label{fig:sAt}
(Color online)
The time dependence of chiral magnetic conductivity $\sA(t)$ (red),
the peak of magnetic helcity spectrum $\kp(t)$ (blue) and magnetic
helicity $h_{m}(t)$ (green).
The schematic divisions of three stages (see text) for the evolution of the system are
sketched in dotted horizontal lines. 
Dashed black curve illustrate $t^{-1/2}$ asymptotic behavior of $\sA(t), \kp(t)$.
Black dotted curve below the green curve plots $h_{m}(t)$ by solving Maxwell's equation in the absence of anomaly effect. 
The numerical results are determined by solving \eq\eqref{master_equation} and \eq\eqref{sA2} with 
$\(L_{\EM} k_{h}, h_{m, I}/h_{0}, \s/\kh\)=\(1/2, 0.8,0.4\)$. 
 }
\end{figure}

\begin{figure}
\centering
       \subfigure[]{
        \includegraphics[width=0.45\textwidth]{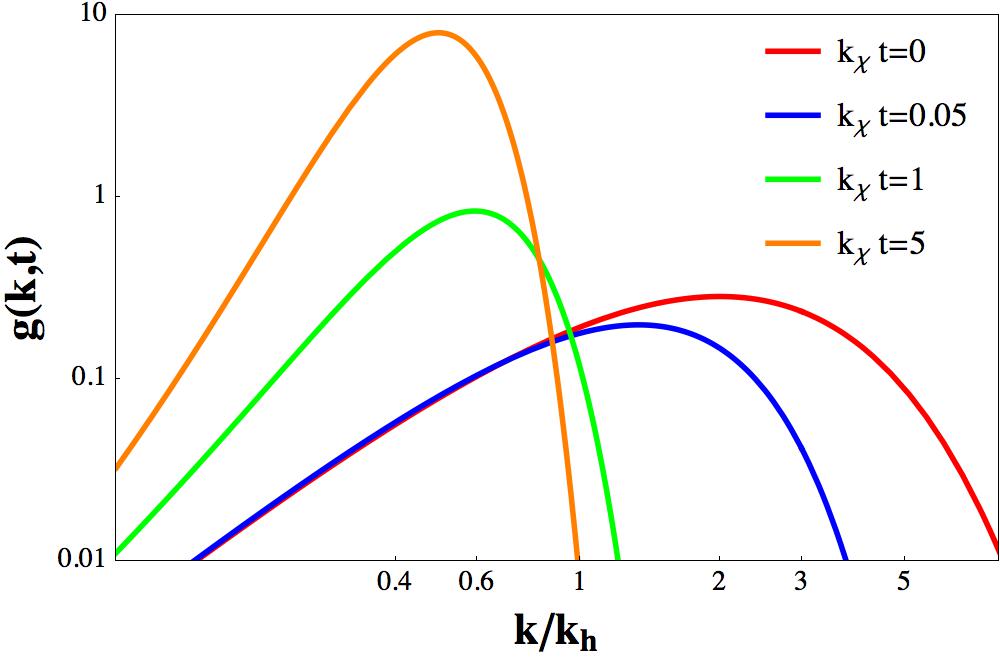}
         \label{fig:gstage}
      }
       \subfigure[]{
        \includegraphics[width=0.45\textwidth]{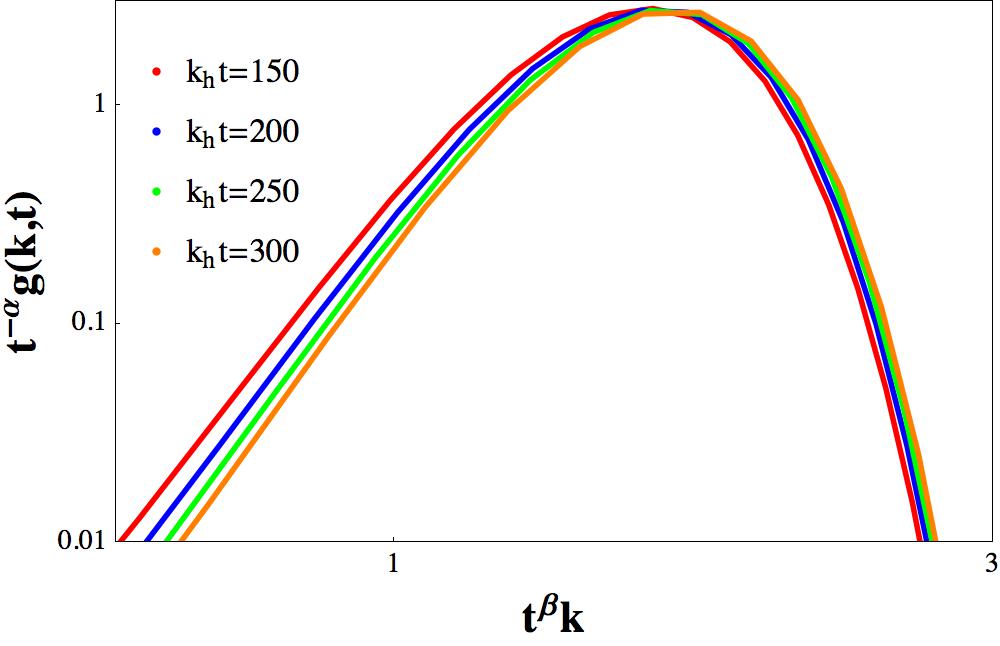}
                \label{fig:gsimilar}
      }
\caption{
\label{fig:bevol}
(Color online) The evolution of magnetic helicity spectrum $g(k,t)$.
(Left): $g(k,t)$ at initial time $t=0$ (red) and three representative
time (corresponding to stage I, II, III) respectively.
(Right): $t^{-\a}g(k,t)$ vs $t^{\b}k$ in the self-similar stage (stage III) of the
evolution. 
}
\end{figure}

In Fig.~\ref{fig:sAt} , 
we plot the time dependence of the chiral magnetic conductivity $\sA(t)$, the peak of $g(\argutk)$,  $\kp(t)$, and magnetic helicity $h_{m}(t)$. 
From Fig.~\ref{fig:sAt}, 
we observe that according to the behavior of $\sA(t), \kp(t), h_{m}(t)$,
 the evolution of system can be generally divided into the following three stages listed below.
For reference,
we also plot the magnetic helicity spectrum $g(\argutk)$ at initial time $t=0$, 
and three representative times corresponding to the three stages in Fig.~\ref{fig:gstage}.
\begin{enumerate}
\item{Stage I:
in this stage, the 
magnetic helicity $h_{m}(t)$ 
decays exponentially.
Due to the conservation of total helicity,
the magnetic helicity is transferred to fermionic helicity -- thus we observe a fast growth of $\sA(t)$.
Meanwhile, $\kp(t)$ starts decreasing but is still larger than $\s_{A}(t)$. 
Stage I ends when magnetic helicity $h_{m}(t)$ becomes small and $\sA(t)$ is close to $\kh$.
The duration of ``stage I'', which we denote as $\tau_{I}$,
can be estimated from the decay rate of magnetic helicity in this stage, 
$\s^{-1}L^{2}_{\EM}$, as indicated by \eq\eqref{g_scaling}.
We therefore have:
\be
\label{tauI}
\tau_{I}
\sim \s L^{2}_{\EM}\, . 
\ee
}
\item{Stage II: 
in this stage, the total helicity $h_{0}$ is dominated by fermionic helicity $h_{F}$. 
In other words, $\sA(t)$ approximately equals to $\kh$ and we observe from Fig.~\ref{fig:sAt} that $\sA(t)$ changes slowly, while 
$\kp(t)$ continues to decrease. 
``Stage II'' ends when $\k_{\rm peak}(t)$ is close to $\sA$. 
}
\item{Stage III:
in this stage, both $\sA(t)$ and $\kp(t)$ decrease. 
The fermionic helicity is transferred to magnetic helicity and eventually 
$h_{m}(t)$ will approach $h_{0}$.
At late times, 
$\sA(t)\approx k_{\rm peak}(t)$.
This corresponds to the configuration in which the energy cost per helicity for fermionic helicity is approximately equal to that of magnetic helicity.  
In this case, 
the following relation holds $\nabla \times \vB \approx k_{\rm peak}(t)\vB \approx \s_{A}(t)\vB$.
It is clear from the log-log plot Fig.~\ref{fig:sAt} that $\sA(t),\kp(t)$ behave as a power law in time 
$t$:
\be
\label{scaling1}
\kh(t)\approx \sA(t) \propto t^{-\beta} \, . 
\ee
Meanwhile, 
the evolution of $g(\argutk)$ becomes self-similar:
\be
\label{g_scaling}
g(\argutk)\sim
t^{\a}\tilde{g}(t^{\beta}k)\, ,
\ee
where $\tilde{g}\(t^{\beta}k\)$ is the scaling function and 
\be
\label{exponent}
\a =1\, , 
\qquad
\b = 1/2\, ,
\ee
are scaling exponents. 
It is easy to see that once $g(\argutk)$ becomes self-similar as in \eq\eqref{g_scaling}, 
$\kp(t)$ is determined by the peak of the scaling function $\tg(t^{\b}k)$.
Therefore $\b$ in \eq\eqref{g_scaling} is identical to $\b$ in \eq\eqref{scaling1}. 

In Fig.~\ref{fig:gsimilar} ,
we plot $g(\argutk)/t^{\a}$ vs $t^{\beta}k$ with scaling exponents given by  \eq\eqref{exponent} for different $t$ in Stage III. 
The self-similar behavior of $g(\argutk)$ is evident from Fig.~\ref{fig:gsimilar}.
At this point, 
critical exponents \eqref{exponent} are found numerically. 
In Sec.~\ref{sec:similar}, we will determine the scaling function $\tilde{g}$ and derive \eqref{exponent} analytically.
}
\end{enumerate}

To close this section,
we emphasize that chiral anomaly plays a crucial role during the evolution of chiral magnetic conductivity $\sA(t)$ and of the magnetic helicity spectrum $g(\argutk)$. 
Indeed, 
without the CME current term in Maxwell equation \eq\eqref{maxwell1} and with no transfer of helicity between magnetic field and chiral fermions, 
$g(\argutk)$ would simple decay as $\exp(-2\s^{-1}k^2t)$ (see also black dotted curve in Fig.~\ref{fig:sAt}) and self-similar behavior would be absent. 

\subsection{Self-similar evolution and scaling behavior of
  $g(\argutk)$
\label{sec:similar}
}
In this section,
we would like to understand the origin of the scaling exponents \eqref{exponent} found
numerically, and to determine the scaling function $\tg(t^{\b}k)$ . 

First of all, we note that  $\a,\b$ are not independent. 
As $h_{m}\approx h_{0}$ at late time (c.f.~Fig.~\ref{fig:sAt}), 
we have:
\be
\label{hm_late}
h_{m}(t)=\int \frac{dk}{\pi} k g(k) 
=\int \frac{dk}{\pi}\, k\, t^{\a}\tg(t^{\b}k) 
= t^{\a-2\beta} 
\int \frac{dx}{\pi}\,  \tg(x)
\approx h_{0}\, ,
\ee
where we have introduced a new variable:
$x\equiv t^{\b}k$. 
We therefore have:
\be
\a=2\b\, . 
\ee

Moreover, if \eq\eqref{g_sol} can be matched to the scaling form
\eq\eqref{g_scaling},
we must have:
\be
\b =\frac{1}{2}\, , 
\qquad
\theta(t) \sim t^{\beta}=t^{1/2}\, . 
\ee
Consequently, 
for self-similar evolution,
we have from \eq\eqref{theta_def}:
\be
\label{theta_s}
\theta(t) = 2\sA(t) t\, . 
\ee
Substituting \eq\eqref{theta_s} into \eq\eqref{g_sol},
we obtain:
\begin{eqnarray}
\label{g_similar}
g(\argutk)
&=& g_{I}(k) \exp\Bigg\{2\s^{-1}\[-k^2 +2\sA(t)\]t\Bigg\}
\nonumber \\
&=&g_{I}(k)\ \exp{\left(-\frac{2\sA^{2}(t)t}{\s}\right)} 
\exp\Bigg\{-2\s^{-1}\[k-\sA(t)\]^2t\Bigg\}\, . 
\end{eqnarray}
If the width of the Gaussian in \eq\eqref{g_similar} is sufficiently narrow,
we further have $\kp(t)\approx\sA(t)$ and 
\be
\label{g_similar2}
g(k,t)
\propto g_{I}(\kp(t))
\exp\Bigg\{-2\s^{-1}\[k-\kp(t)\]^2t\Bigg\}\, . 
\ee

To summarize,
the system will spend a long time at the stage of self-similar evolution. 
In this stage, $\kp(t)$ decreases as $t^{-1/2}$.
This implies that a large-scale helical magnetic field will develop.
With the growth of $t$,
the width of the Gaussian becomes more and more narrow, and $g(\argutk)$ will become proportional to delta-function :
\begin{equation}
g(\argutk)\to \delta\[k-\kp(t)\]\, . 
\end{equation}
In this limit, 
the system is described by a single CK state $W^{+}_{11}(\kp(t),t)$. 
Eventually, the evolution will end when $\kp(t)\sim 1/L$ where $L$ is the size of the system.
Here we have found self-similar evolution by solving the MCS equation with the Hopfion initial condition \eq\eqref{Hopf_sol}. 
However, 
as our analysis does not rely on any particular feature of the Hopfion solution,  
we expect that self-similar evolution is a general feature that at late times does not depend on the choice of initial conditions.

\section{Implications for heavy-ion collisions
\label{sec:HIC}}

Let us now establish whether a 
 self-similar
evolution of magnetic helicity can be realized in experiment;  here we will focus on QCD matter created in heavy-ion collisions. 
As we discussed in Sec.~\ref{sec:energy},
to realize the inverse cascade of magnetic helicity,
the energy cost per helicity  if total helicity is carried by chiral fermions $k_{h}$ (see \eq\eqref{kh_def}),
should be larger than $k_{\rm min}$,
the minimum eigenvalue of a CK state. 
As $k_{\rm min}$ is of the order $1/L$, where $L$ is the size of the system,
we need to check whether $k_{h}>1/L$ with $L\sim 10~\fm$ can be realized in a heavy-ion collision.
As both magnetic helicity and fermionic helicity would contribute to the
total helicity,
we will estimate their contributions separately. 
In heavy-ion collisions,
the initial axial charge density can be generated by sphaleron transitions
and/or by the color flux tubes during the early moments of the heavy-ion
collision~\cite{Hirono:2014oda}.
To make a rough estimate, we will follow Ref.~\cite{McLerran:2013hla} and consider an (optimistically large) value of the chiral chemical potential $\muA$, of the order of $1$~GeV.
The resulting $\sA$ is then of the order $0.01~{\rm GeV}$. 
If the total helicity $h_{0}$ originates mostly from this initial axial charge,
$1/k_{h}$ would be at least of order $20~{\rm fm}$,
which is much larger than the typical size of the fireball $L\sim 10~\fm$ created in
a heavy-ion collision. 
This estimate is also in agreement with Ref.~\cite{Chernodub:2010ye} in which the relevance of CK state to heavy-ion collisions was discussed. 
To summarize,
 in order to satisfy $\kh>1/L$, 
initially the dominant contribution to the total helicity should be from magnetic helcity $h_{m}$. 
\begin{figure}
\centering
        \includegraphics[width=0.6\textwidth]{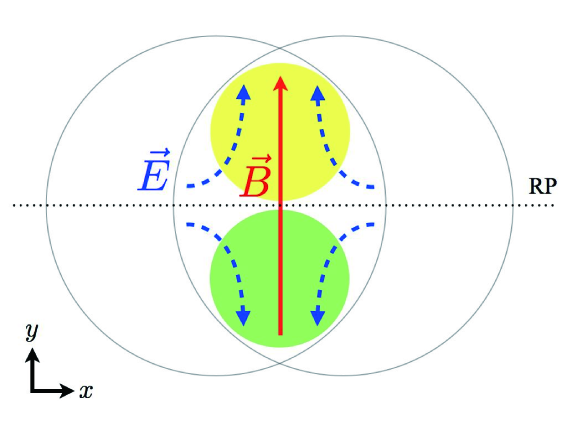}
       \caption{
  \label{fig:EB_HIC}
(Color online) 
A typical configuration of EM field in the transverse plane in non-central collisions. 
The two circles indicate the edge of the colliding nuclei. 
The solid line (red) shows the magnetic field $\vB$ and the dashed lines (blue) show the electric field $\vE$. 
The inner product between electric and magnetic fields $\vE\cdot\vB⃗$ becomes positive (negative) on the upper (lower) side of reaction plane (RP).
}
\end{figure}
\vskip0.3cm

We now estimate magnetic helicity in a heavy-ion experiment. 
We first note that from the EM field pattern created by spectators in heavy-ion collisions  (c.f.~Fig~\ref{fig:EB_HIC}),
 $\vE\cdot \vB$ is positive in the upper half region, and negative in the lower half region. 
The EM field would thus be helical, with opposite helicities in the upper and lower half region. 
To estimate the magnitude of this magnetic helicity,
 we assume $|\vB|\approx |\vE|$ and in RHIC $\vA \sim \vE\ \tau_{B, \RHIC}$ where $\tau_{B,RHIC}$ is the typical lifetime of magnetic field at RHIC. 
The typical (peak) strength of magnetic field at RHIC is
\be
eB_{\RHIC} = c_{B} m^2_{\pi}\, , 
\ee
where $c_{B}$ varies in the range of  $1$ to $10$
depending on the impact parameter after event averaging, but in a given event can be significantly larger than this average value due to fluctuations.
We therefore have:
\begin{eqnarray}
k_{h,\RHIC} 
&\approx&
\frac{C^2_{A} \int d^3x \vA_{\RHIC}\cdot\vB_{\RHIC}}{\chi_{A}V}
\approx\frac{e^{-2}C^2_{A}\(eB_{\RHIC}\)^2 \tau_{B,RHIC}}{N_f \chi_{f}}
\nonumber\\
&=& \a_{\EM}\(C_{\EM}\)^2\(\frac{N^2_{c}}{4\pi^3}\)\,
          \frac{\(eB_{RHIC}\)^2\tau_{B,RHIC}}{N_{f}\chi_{f}}
=1.0\times 10^{-5} c^2_{B}\,  \tau_{B,RHIC}~\fm^{-2}\, . 
\end{eqnarray}
Here we have assumed that the axial susceptibility is 
$\chi_{A}=N_{f}\chi_{f}$ where $\chi_{f}$ is the quark number 
susceptibility known from the lattice measurements~\cite{Borsanyi:2011sw,Bazavov:2012jq} to be $\chi_{f}\approx 1.0
T^2$ for temperatures higher than $T_{c}$. 
We consider the case of $u,d$ flavors contributing to the CME current
and thus put $N_{f}=2, C_{\EM}=5/9$. 
We also assume that at the initial stage of heavy-ion collisions $T\approx 0.4~{\rm GeV}$.  
To compute $k_{h}$ at LHC,
we further take:
\be
\frac{eB_{\LHC}}{eB_{\RHIC}}\, ,
\approx 
\frac{\g_{\LHC}}{\g_{RHIC}}
\approx 13.8\, \qquad
\frac{\tau_{B, \LHC}}{\tau_{B, \RHIC}}
\approx 
\(\frac{\g_{\LHC}}{\g_{RHIC}}\)^{-1}
.
\ee
We then get:
\be
k_{h,\LHC}
\approx 
\(\frac{\g_{\LHC}}{\g_{\RHIC}}\) k_{h,\RHIC}
= 1.4\times 10^{-4}c^2_{B} \tau_{B,\RHIC}~\fm^{-2}\, . 
\ee
For the purpose of estimate,
we will take $\tau_{B, \RHIC}=1~{\rm fm}$.
Therefore in order to satisfy $k_{h}>1/L\approx 0.1~{\rm fm}^{-1}$,
we need to select events with $c_{B}\sim 100$ at RHIC and $c_{B}\sim 26$ at LHC.  
\vskip0.3cm

We conclude that observing the self-similar cascade of magnetic helicity in heavy ion collisions will be challenging. However the estimated magnetic helicity is quite large, and is likely to affect the evolution of the quark-gluon plasma. It can lead to interesting observable effects. For example, since the sign of magnetic helicity is different in the upper and lower hemispheres 
(above and below the reaction plane), the decay of the magnetic field at freeze-out will yield the photons with opposite circular polarizations. 
Since the direction of magnetic field, and thus the signs of magnetic helicity, can be determined experimentally by measuring the spectators, one can 
measure the polarizations of photons by summing over many events (the sign of polarization will not fluctuate event-by-event). 
The photon polarization can be measured through photon
conversion into $e^{+} e^{-}$ pairs by extracting the angular distribution of the electrons and positrons.
We believe that the observation of these opposite circular polarizations of the produced photons will be a unique signature of magnetic helicity in heavy ion collisions.  
 
\section{Summary
\label{sec:conclusion}
}

To summarize, the chiral anomaly couples the evolution of axial charge density and electric-magnetic (EM) field in the plasmas possessing chiral fermions. 
By solving the Maxwell-Chern-Simons equation in the presence of CME current, 
we analyzed  the real time evolution of the magnetic helicity spectrum. 
We initialized the system by assuming that it contains a seed of a helical magnetic field at short distances, with helicity carried by the linked magnetic field configuration, i.e. by the second term in (\ref{mag_top}). 
As summarized in Fig.~\ref{fig:sAt}, we found that 
the magnetic helicity first gets transferred to fermionic helicity and then fermionic helicity is transformed back into magnetic helicity, but this time to self-linked Chandrasekhar-Kendall (CK) configurations characterized by the second term in (\ref{mag_top}). We have argued that the CK states that minimize magnetic energy at a fixed helicity represent the final stage of the magnetic helicity evolution.  
We found that at late stage, 
this evolution becomes self-similar, and describes the growth of the self-linked CK state. 
\vskip0.3cm

The role of fermions is to mediate the magnetic reconnections that are necessary to transfer the magnetic helicity from the linked to self-linked configurations of magnetic flux, see (\ref{mag_top}). 
We expect that our findings apply to all systems that possess the CME current. In addition to the quark-gluon plasma discussed above, the growth of magnetic helicity can be expected in Dirac semimetals that exhibit the CME in parallel electric and magnetic fields \cite{Li:2014bha}. Experimentally, this generation of magnetic helicity can manifest itself through the emission of circularly polarized photons in the THz frequency range characteristic for Dirac semimetals \cite{Kharzeev:2014sba}. 
\vskip0.3cm

As a natural extension of this work, it will be interesting to study the inverse cascade in the framework of anomalous MagnetoHydroDynamics (MHD). 
In this case, the anomaly can couple the kinetic helicity carried by the fluid, magnetic helicity and fermionic helicity. While the inverse cascade of magnetic helicity is a traditional topic of magnetohydrodynamics, the role played by the chiral anomaly has not yet been fully explored.  As another extension of this work, it would be interesting to include the spatial dependence of axial charge density in the MCS equations.
This would allow us to study the evolution of domains with non-zero axial charge throughout the inverse cascade of magnetic helicity.  

\acknowledgments
We thank Larry McLerran, Soren Schlichting, Paul Wiegmann, Aihong Tang and Ho-Ung Yee for useful discussions,
Roman Jackiw for helpful comments on the manuscript
and Carlos Hoyos for communication on the spectral representation of the Hopfion solution. 
This work was supported by the U.S. Department of Energy under Contracts
No. DE-FG-88ER40388 (D.K.) and by DOE grant No. DE-SC0012704 (D.K and Y.Y.) . 
Y. H. is supported by the JSPS Research Fellowship for Young Scientists.

\begin{appendix}
\section{A useful representation of Chardrasekhar-Kendall states
\label{sec:CK}}
A single CK state $\vW^{\pm}(\vx;k)$ can be represented as the linear combination of toroidal field $\vT(\vx;k)$ and
poloidal field $\vP(\vx;k)$ 
\be
\vW^{\pm}_{lm}(\vx;k) = \vT_{lm}(\vx; k)\mp i \vP_{lm}(\vx;k)\, , 
\ee
where 
\be
\vT_{lm}(\vx;k)= j_{l}(k r)\vX_{lm}(\theta, \phi)\, , 
\qquad
\vP_{lm}(\vx;k)= \frac{i}{k}\nabla \times \vT_{lm}(\vx;k)\, ,
\ee
where $j_{l}(kr)$ denotes spherical Bessel functions. 
Here, we have defined:
\be
\vX_{lm}(\theta,\phi)
\equiv \frac{1}{\sqrt{l(l+1)}}\vL \[Y_{lm}(\theta,\phi)\]\, , 
\qquad
\vL\equiv- i \(\vr\times \nabla\)\, . 
\ee
\end{appendix}
In the above equations, $Y_{lm}(\theta,\phi)$ is the usual speherical harmonic functions.

\end{document}